\documentclass[preprint,superscriptaddress,amsmath,amssymb,prd,aps,showpacs,floatfix,nofootinbib]{revtex4-1}
\usepackage{graphicx}
\usepackage{dcolumn}
\usepackage{bm}
\usepackage{mathrsfs}
\usepackage{hyperref}
\usepackage{amsmath}
\usepackage{amssymb}
\usepackage{bm}
\usepackage{color}
\usepackage{float}
\usepackage{dcolumn}
\usepackage{multirow}
\usepackage{changepage}
\hypersetup{pdftex,colorlinks=true,linkcolor=blue,citecolor=red,menucolor=black,urlcolor=blue,filecolor=blue}
\newcommand{\tabincell}[2]{\begin{tabular}{@{}#1@{}}#2\end{tabular}}

\begin{document}
\title{Observed $\Omega_b$ spectrum and meson-baryon molecular states}
\date{\today}

\author{Wei-Hong~Liang}
\email{liangwh@gxnu.edu.cn}
\affiliation{Department of Physics, Guangxi Normal University, Guilin 541004, China}
\affiliation{Guangxi Key Laboratory of Nuclear Physics and Technology, Guangxi Normal University, Guilin 541004, China}

\author{E.~Oset}
\email{oset@ific.uv.es}
\affiliation{Departamento de F\'{\i}sica Te\'orica and IFIC, Centro Mixto Universidad de
Valencia-CSIC Institutos de Investigaci\'on de Paterna, Aptdo.22085,
46071 Valencia, Spain}
\affiliation{Department of Physics, Guangxi Normal University, Guilin 541004, China}


\begin{abstract}
We observe that four peaks seen in the high energy part of the $\Omega_b$ spectrum of the recent LHCb experiment
are in remarkable agreement with predictions made for molecular $\Omega_b$ states stemming from the meson-baryon interaction,
with an approach that applied to the $\Omega_c$ states gave rise to three states in good agreement with experiment in masses and widths.
While the statistical significance of the peaks prevents us from claims of states at the present time,
the agreement found should be an incentive to look at this experiment with increased statistics to give an answer to this suggestive idea.
\end{abstract}

\maketitle


\section{Introduction}
\label{sec:intro}

The recent LHCb experiment on the $\Omega_b$ excited spectrum has reported four new peaks that have been associated with new excited $\Omega_b$ states,
$\Omega_b(6316)$, $\Omega_b(6330)$, $\Omega_b(6340)$, and $\Omega_b(6350)$ \cite{LHCb}.
The experiment has already stimulated theoretical work and in Ref.~\cite{Hosaka}, using QCD sum rules,
the suggestion is made that the states correspond to ordinary $1P$ excitations of three quark states.
Similar claims are made in Ref.~\cite{Qifang} where decay modes of these states are discussed.

In this short paper we discuss that these structures do not correspond to molecular states originated from the meson-baryon interaction.
However, we call attention to four other peaks observed at higher energy,
albeit with low statistics,
which remarkably agree with predictions made with the molecular picture \cite{Omegab}.
These peaks occur at $6402$, $6427$, $6468$, and $6495$ MeV, and the statistical significance is equivalent to that of the peak at $6315.64$ MeV,
which has given name to one of the states claimed, the $\Omega_b(6316)$.
In Ref.~\cite{Omegab} four states were found at $6405~ {\rm MeV} (\frac{1}{2}^-)$, $6427~ {\rm MeV} (\frac{3}{2}^-)$, $6465~ {\rm MeV} (\frac{1}{2}^-)$, and
$6508~ {\rm MeV} (\frac{1}{2}^-, \frac{3}{2}^-)$,
which well match the position of the peaks reported above.
The widths of these states are smaller than $1$~MeV,
except for the $6465$~MeV state that has a width of $2.4$ MeV.

There is abundant work done on bottom baryons in the literature, mostly in quark models.
A relativized quark model was earlier used in Ref.~\cite{10chen},
and the recent progress detecting many such states has motivated much theoretical work recently.
Nonrelativistic quark models are used in Refs.~\cite{12chen,13chen,14chen,Santopinto},
full calculations with Faddeev equations are done in Ref.~\cite{Vijande},
a relativistic quark model is used in Refs.~\cite{Faustov,Ebert,15chen},
heavy quark effective theory is used in Ref.~\cite{roperuin},
the quark pair creation model is used in Refs.~\cite{20chen,22chen,Qifang},
the relativistic flux tube model is used in Ref.~\cite{24chen},
the color hyperfine interaction is used in Refs.~\cite{25chen, 26chen},
the chiral quark model is used in Refs.~\cite{16chen,17chen,18chen},
and chiral perturbation theory is used in Ref.~\cite{27chen}.
QCD sum rules have also been much used to calculate the $\Omega_b$ spectrum
and decay widths \cite{ChenZhu,ChenHosaka,ChenLiu,Narison,Azizi,Agaev,HXChen,49chen,52chen},
and as usual, they have uncertainties in the masses of the order of 100 MeV or more.
Lattice QCD calculations have also brought their share to this topic \cite{mathur}.
The QCD motivated hypercentral quark mode is used in Refs.~\cite{Thakkar,Shah}.
Some reviews on this issue are given in Refs.~\cite{43chen,44chen,45chen,46chen,47chen,48chen,25chen}
and an update of more recent works can be seen in Refs.~\cite{chenrefs,Qifang}.

Most of the works on quark models calculate the mass spectra of the states but do not evaluate decay widths.
There are, however, some works devoted to this task,
mostly using the $^3P_0$ model \cite{14chen,21chen,Santopinto,Qifang}
or the chiral quark model \cite{qzhao,16chen,18chen,34lu}.

By contrast to the rich spectra of states in the quark model, the molecular states of $\Omega_b$ nature are much fewer,
and rather than $500$ MeV quark excitation in the quark configurations,
one is talking about a few MeV binding of the main components.
Provided one can tune the few free parameters to some experimental data,
the predictions are relatively accurate.
In the present case, such relevant information is provided by the analogous $\Omega_c$ states reported in Ref.~\cite{ExpOmegac}.

In the next section, we discuss the origin of the masses for the molecular $\Omega_b$ states in connection with the spectrum of the related $\Omega_c$ states.

\section{$\Omega_c$ and $\Omega_b$ molecular states}
\label{sec:generate}

It is very interesting to compare the spectrum of the $\Omega_c$ states reported by the LHCb \cite{ExpOmegac} with that of the $\Omega_b$ states.
In Ref.~\cite{ExpOmegac} five narrow states were observed and named $\Omega_c (3000)$, $\Omega_c (3050)$,
$\Omega_c (3066)$, $\Omega_c (3090)$, and $\Omega_c (3119)$.
Three of these states were reproduced in Ref.~\cite{Omegac},
both in the mass and the width,
the $\Omega_c (3050)$, $\Omega_c (3090)$, and $\Omega_c (3119)$ states.
The states were obtained by solving the Bethe-Salpeter equation in the coupled channels
$\Xi_c\bar{K}$, $\Xi^{\prime}_c\bar{K}$, $\Xi D$, $\Omega_c \eta$, $\Xi D^*$, $\Xi_c \bar{K}^*$, $\Xi^{\prime}_c\bar{K}^*$,
$\Xi^*_c \bar{K}$,  $\Omega^*_c \eta$, and $\Xi^* D$,
with an interaction based on the exchange of vector mesons
extending the local hidden gauge approach \cite{hidden1,hidden2,hidden4,hidden3,hideKo} to the charm sector.
The meson-baryon loop functions were regularized using the cutoff method,
with $q_{\rm max}=650$ MeV,
the maximum value of the modulus of the three-momentum in the loop.

The scattering matrix for transitions between channels is given by
\begin{equation}\label{eq:BSeq}
T = [1 - V G]^{-1}\, V ,
\end{equation}
with $V$ the transition potential which is found as
\begin{eqnarray} \label{eq:Vij}
V_{ij}=D_{ij} \frac{1}{4f^2}(p^0+p^{\prime \,0}) ,
\end{eqnarray}
with $f$ the pion decay constant $f=93$ MeV,
$p^0$, $p^{\prime 0}$ the energies of the initial and final mesons, respectively,
and $D_{ij}$ coefficients which are evaluated in Ref.~\cite{Omegac}.
The couplings of the states obtained and the wave function at the origin were evaluated,
which allowed us to identify the most important coupled channel for each state.

In Table \ref{tab:tab1} we show the results reported in Ref.~\cite{Omegac},
together with the main channel for each state,
its threshold mass and the diagonal $D_{ij}$ coefficient for this channel.
\begin{table*}[tb]
\renewcommand\arraystretch{1.5}
\centering
\caption{\vadjust{\vspace{-0pt}}Experimental $\Omega_c$ states from Ref.~\cite{ExpOmegac} and theoretical predictions from Ref.~\cite{Omegac}.
The main channel for each state is shown together with its threshold mass and the diagonal $D_{ii}$ coefficient for this channel.
In brackets is the width of the state.
Mass units are in MeV.}
\label{tab:tab1}
\begin{tabular*}{0.96\textwidth}{@{\extracolsep{\fill}}c|cccc}
\hline
\hline
Experimental state & \tabincell{c}{$\Omega_c (3050)$ \\[-0.5cm]  $[0.8 \pm 0.2 \pm 0.1]$} & \tabincell{c}{$\Omega_c (3090)$ \\[-0.5cm] $[8.7 \pm 1.0\pm 0.8]$} & \tabincell{c}{$\Omega_c (3119)$ \\[-0.5cm]$[1.1\pm 0.8 \pm 0.4]$} & \\
\hline
Ref.~\cite{Omegac} & $3054~(\frac{1}{2}^-)\; [0.88]$ & $3091~ (\frac{1}{2}^-)\; [10.2]$ & $3125~ (\frac{3}{2}^-)\; [0]$
& $3221~ (\frac{1}{2}^-, \frac{3}{2}^-)\; [0]$ \\
 Main channel & $\Xi^{\prime}_c\bar{K}$ & $\Xi D$ &$\Xi_c^* \bar K$ & $\Xi D^*$ \\
 Threshold mass & $3074$ & $3185$  & $3142$ & $3327$\\
 $D_{ii}$ & $-1$ & $-2$  & $-1$ & $-2$\\
 \hline\hline
\end{tabular*}
\end{table*}
We can see that $\Omega_c (3050)$ is associated with a $J^P=\frac{1}{2}^-$ state at $3054$ MeV,
which is mostly a $\Xi^{\prime}_c\bar{K}$ state bound by about $20$ MeV.
The other $\frac{1}{2}^-$ state at $3091$ MeV is associated with the $\Omega_c (3090)$.
It is mostly a $\Xi D$ state bound by about $94$ MeV.
Finally, we have a $\frac{3}{2}^-$ state at $3125$ MeV, which we associate with $\Omega_c (3119)$.
It is mostly built up from the $\Xi^*_c \bar{K}$ channel and is bound by about $17$ MeV.
In addition, we also put in Table \ref{tab:tab1} a state found at $3221$ MeV,
degenerate in $\frac{1}{2}^-, \frac{3}{2}^-$, which couples mostly to the $\Xi D^*$ component and is bound by $106$ MeV.
We can see that the $3054$ and $3125$ MeV states are moderately bound by about $20$ MeV,
while the other two states are bound by about $100$ MeV.
The reason for this difference is the strength of the interaction represented by the $D_{ij}$ coefficient of Eq.~\eqref{eq:Vij},
which is double the other for the two most bound states.

The two $\frac{1}{2}^-$ states of Table \ref{tab:tab1} were earlier obtained in Ref.~\cite{Montana},
with basically the same mass and width and the same main component,
using an extension of the chiral unitary approach, with an interaction as the one in Eq.~\eqref{eq:Vij}
which gave the same coefficients for the diagonal transitions.
The $\frac{3}{2}^-$ state was not obtained because the $\Xi^*_c$ baryon was not considered in the coupled channels.
A different approach was used in Ref.~\cite{PavaoJuan} using an ${\rm SU(6)}_{lsf} \times {\rm SU(2)}$ heavy quark spin symmetry (HQSS) extension of the Weinberg-Tomozawa interaction,
combining SU(6) light quark spin-flavor symmetry and heavy quark spin symmetry for the heavy quarks.
This approach leads to similar qualitative features,
but depending on the regularization,
it gives rise to more bound states.
An approach based on meson exchange is used in Ref.~\cite{ChenHosa}
where a $\frac{3}{2}^-$ $\Omega_c$ bound state is found that couples mostly to $\Xi^*_c \bar{K}$, as in our case.
Other works assuming molecular nature for the $\Omega_c$ states are Refs.~\cite{He,Huang,Pavon}, but there, only the $\Xi D$
( also $\Xi D^*$ in Ref.~\cite{Pavon}) components are considered,
and the decay modes are studied.
The results in Table \ref{tab:tab1} are obtained with all coupled channels,
and using an interaction which has proved very successful reproducing the LHCb pentaquark states \cite{pentaEx} in Ref.~\cite{XiaoJuan}.

One could apply arguments of HQSS \cite{Isgur,Neubert,manohar}
to relate the states of Table \ref{tab:tab1} with their analog ones of $\Omega_b$,
changing a quark $c$ by the quark $b$.
Instead of this, we study the molecular $\Omega_b$ states in Ref.~\cite{Omegab}
in the same footing as the $\Omega_c$ states, exchanging vector mesons to generate the interaction that is used together with the Bethe-Salpeter equation
and the corresponding coupled channels.
Yet, the approach respects HQSS, since in the transitions where light vectors are exchanged the heavy quarks are spectators,
and when heavy vectors are exchanged, the terms are suppressed in the heavy quark mass counting.
This symmetry refers to the $V_{ij}$ transition potentials,
but it was discussed in Ref.~\cite{AltenGeng} that extra caution should be exerted when regularizing the loops,
which in the cutoff regularization method turned into using the
same cutoff in the charmed or bottom sectors \cite{LuGeng,Ozpineci}.

With all these considerations in mind, the $\Omega_b$ states were studied in Ref.~\cite{Omegab} with the coupled channels
$\Xi_b\bar{K}$, $\Xi^{\prime}_b\bar{K}$, $\Omega_b \eta$, $\Xi \bar B$, $\Xi^*_b \bar K$, $\Omega^*_b \eta$, $\Xi^* \bar B$, $\Xi \bar B^*$, $\Xi_b \bar{K}^*$, and $\Xi^{\prime}_b \bar{K}^*$ using the same cutoff ($q_{\rm max}=650~ {\rm MeV}$) as for the $\Omega_c$ states,
and the results are summarized in Table \ref{tab:tab2},
together with the potential experimental states extracted from the peaks at high energy in the $\Omega_b$ spectrum of Ref.~\cite{LHCb}.

\begin{table*}[tb]
\renewcommand\arraystretch{1.5}
\centering
\caption{\vadjust{\vspace{-0pt}} Peak energies from Ref.~\cite{LHCb} at the higher energy region and theoretical predictions from Ref.~\cite{Omegab}. The main channel for each state is shown together with its threshold mass and the diagonal $D_{ii}$ coefficient for this channel. Mass units are in MeV.}
\label{tab:tab2}
\begin{tabular*}{0.9\textwidth}{@{\extracolsep{\fill}}c|cccc}
\hline
\hline
Experimental peaks & $6402$ & $6427$ & $6468$ & $6495$ \\
\hline
Ref.~\cite{Omegab} & $6405~(\frac{1}{2}^-)$ & $6427~ (\frac{3}{2}^-)$ & $6465~ (\frac{1}{2}^-)$ & $6508~ (\frac{1}{2}^-, \frac{3}{2}^-)$ \\
 Main channel & $\Xi^{\prime}_b\bar{K}$ & $\Xi_b^* \bar K$ &$\Xi \bar B$ & $\Xi \bar B^*$ \\
 Threshold mass & $6431$ & $6451$  & $6598$ & $6643$\\
 $D_{ii}$ & $-1$ & $-1$  & $-2$ & $-2$\\
 \hline\hline
\end{tabular*}
\end{table*}

Unlike the case of the $\Omega_c$, there is not much work on molecular $\Omega_b$ states.
One exception is the work of Ref.~\cite{Tolos} where, using ${\rm SU(6)}_{lsf} \times {\rm SU(2)}$ HQSS symmetry,
a sextet is obtained with which the $\Sigma_b (6097)$ and $\Xi_b(6227)$ states are associated and extrapolating
the masses predicts an $\Omega_b$ state at $6360$ MeV.

The features of the states obtained are nearly a copy of what is found for the $\Omega_c$ states.
We find a $\frac{1}{2}^-$ state at $6405$ MeV which couples mostly to $\Xi^{\prime}_b\bar{K}$ and is bound by about $26$ MeV.
Another state of $\frac{3}{2}^-$ at $64427$ MeV couples mostly to $\Xi_b^* \bar K$ and is bound by about $24$ MeV.
On the other hand, the $\frac{1}{2}^-$ state at $6465$ MeV couples mostly to $\Xi \bar B$ and is bound by about $133$ MeV,
while the $\frac{1}{2}^-, \frac{3}{2}^-$ state at $6508$ MeV couples mostly to $\Xi \bar B^*$ and is bound by about $135$ MeV.
The main channels of the states found are the same as in Table \ref{tab:tab1} changing a $c$ quark by a $b$ quark and the binding energies are very similar.
We should note that the order of energies of the $\Xi_b^* \bar K~ (\frac{3}{2}^-)$, $\Xi \bar B~ (\frac{1}{2}^-)$
and $\Xi_c^* \bar K~ (\frac{3}{2}^-)$, $\Xi D~ (\frac{1}{2}^-)$ states is reversed,
but this is not due to different bindings but to the masses of the components.
The bindings are similar in both cases.
Once again we see that two of the states are moderately bound and they correspond to a diagonal interaction with $D_{ii}=-1$,
while the other two states, which are more bound, correspond to $D_{ii}=-2$,
with double strength in the interaction.

As to the width of the states that we get, we have $\Gamma\simeq 0$ MeV for the $6405\, {\rm MeV} \, (\frac{1}{2}^-)$, $6427\, {\rm MeV} \, (\frac{3}{2}^-)$, and $6508\, {\rm MeV} \, (\frac{1}{2}^-,\, \frac{3}{2}^-)$ states,
and $\Gamma\simeq 2.4$ MeV for the $6465\, {\rm MeV} \, (\frac{1}{2}^-)$ state.
The peaks that we refer to in Ref.~\cite{LHCb} are very narrow,
and by comparison to the $\Omega_b(6316)$, which has a width $\Gamma_{\Omega_b}< 2.8$ MeV \cite{LHCb} we can reasonably guess that should these peaks be confirmed as states in future runs, the width is also smaller than $2.8$ MeV,
which would be in agreement with our findings.

The measurement of the widths is also relevant and could be a decisive factor to distinguish molecular states from ordinary $3q$ states.
In this sense it is worth recalling that for the $\Omega_c$ molecular states associated with the experimental ones,
the agreement of the widths with experiment was remarkable,
as one can see in Table \ref{tab:tab1}.
In this sense, it is worth recalling the result of $\Gamma_{\Omega_b (3/2^-)}=58^{+65}_{-33}$ MeV of Ref.~\cite{HXChen} for a state at $6460$ MeV
and widths evaluated for other states that have a very wide margin of variation.

There is one more reason that we can provide to argue that the states reported in Ref.~\cite{LHCb} are not of molecular nature.
We can artificially increase $q_{\rm max}$ to match one of the states and see what we obtain for the other ones.
We find that with $q_{\rm max}=830$ MeV, we get a $\frac{1}{2}^-$ state at $6317$ MeV
that matches the $\Omega_b (6316)$ in mass,
but the width becomes $13$ MeV, in disagreement with the width smaller than $2.8$ MeV claimed in Ref.~\cite{LHCb}.
Also, the binding of the $\Xi'_b \bar K$ component would be $115$~MeV,
which we find too big for the strength of the interaction with coefficient $D_{ii}=-1$ compared with the $20$ MeV binding of the equivalent $\Omega_c$ state.
Furthermore, we obtain then for the masses of the other states $6358~{\rm MeV} (\frac{1}{2}^-)$,
$6362~{\rm MeV} (\frac{1}{2}^-, \frac{3}{2}^-)$, and $6375~{\rm MeV} (\frac{3}{2}^-)$, none of which matches the mass of the reported $\Omega_b$ states.

It is clear that the peaks seen in the $\Omega_b$ spectrum reported in Table II do not have enough statistics to make a claim of states,
but we should emphasize that the statistical significance is the same as for the $\Omega_b(6316)$ claimed in Ref.~\cite{LHCb}.
Yet, it is difficult to think that the impressive agreement of the energy of these four peaks with the predictions of Ref.~\cite{Omegab} is a pure coincidence.
This observation should be reason enough to look into the experiment and come back with more statistics to corroborate this ansatz or refute it,
which is the main purpose of the present work.

\section{Comparison with the results of quark models}
  It is interesting to establish a comparison with the results of quark models.
  Most of the recent work on quark models tries to describe the reported $\Omega_b$ states of Ref.~\cite{LHCb}.
  The majority of the works obtain $\Omega_b$ states around $6350$ MeV for the $1P$ excitations with $J^P=\frac{1}{2}^-, \frac{3}{2}^-$.
  The widths depend much on the models.
  In Ref.~\cite{16chen} a width of $143$ MeV is obtained for one of the low-lying $\frac{1}{2}^-$ states while $95$ MeV is obtained in Ref.~\cite{qzhao}.
  For another $\frac{1}{2}^-$ state, Ref.~\cite{qzhao} obtains $49$ MeV, but both these works give widths smaller than $2$ MeV for the $\frac{3}{2}^-$ states.
  Similar values are obtained in Ref.~\cite{18chen} and smaller values in Refs.~\cite{21chen,Qifang}.
  In Ref.~\cite{Qifang}, one of the $\frac{1}{2}^-$ states has, however, a width as large as $1000$ MeV,
  which, as discussed there, would make the state unobservable,
  and even bigger values are found within the QCD sum rule calculation of Ref.~\cite{HXChen}.

The states that we propose appear from $6400$ to $6500$ MeV and are not those reported in Ref.~\cite{LHCb}.
The first lesson one learns is that the first molecular states
that appear in Ref.~\cite{Omegab} with $\frac{1}{2}^-, \frac{3}{2}^-$ are very different from those of the quark model,
which appear from $1P$ excitation roughly about $100$ MeV below those of Ref.~\cite{Omegab}.
Then in the quark models, one has to go to the $2S$ excited states,
which are in the region of $6450$ MeV \cite{roperuin,15chen,Faustov,Santopinto,21chen} to eventually match the energies of the peaks that we propose as states.
Some models predict the masses a little lower, below $6400$ MeV \cite{Vijande,Segovia},
and the QCD sum rules of Ref.~\cite{maohosa} above $6000$ MeV, although compatible with values around $6450$ MeV considering the uncertainties.
Then, these states would have positive parity, unlike the molecular states that have negative parity.
Here, one can see a very distinctive dynamical character for these models,
and measuring the quantum numbers of these states can clearly discriminate between them.

As to the widths of the states, according to Ref.~\cite{21chen} the $2S$ quark model states are expected to have widths of around $15$ MeV,
but there are also differences between the different models.

In our molecular model, one can obtain the widths from the couplings to the different channels.
This is a peculiar feature of the coupled channels approach,
since different orthogonal combinations of the channels are obtained as eigenstates, and the couplings vary much from one state to another.
The case is different for the results that one would obtain from the $^3P_0$ decay model in quark configurations.

In this sense, we can go to Ref.~\cite{Omegab} and, by using the couplings of the states obtained there to the open channels,
we can evaluate the partial decay widths.
We have the decay width given by
   \begin{equation}\label{eq:1here}
     \Gamma=\dfrac{1}{2\pi} \dfrac{M_{\rm fin}}{M_{\rm in}} \,|g|^2 \, p_m,
   \end{equation}
where $M_{\rm fin}$, $M_{\rm in}$ are the mass of the final baryon in the decay channel and the mass of the obtained $\Omega_b$ state,
and $p_m$ is the meson momentum of the meson-baryon decay channel.
We find that the $6405$ MeV state is only above the $\Xi_b \bar K$ threshold and only decays into that channel.
The $6465$ MeV state decays into $\Xi_b \bar K$ and $\Xi'_b \bar K$.
The state at $6427$ MeV couples to $\Xi^*_b \bar K$ and other channels that are all above this energy.
Similarly, the state at $6508$ MeV couples to $\Xi \bar B^*$ and other channels which are all closed.
Using Eq.~\eqref{eq:1here} and the couplings of Ref.~\cite{Omegab} we obtain the partial decay widths of Table \ref{tab:new}.
\begin{table*}[tb]
\renewcommand\arraystretch{1.5}
\centering
\caption{\vadjust{\vspace{-0pt}} Partial decay widths of the predicted $\Omega_b$ states to decay channels (units in MeV).}
\label{tab:new}
\begin{tabular*}{0.6\textwidth}{@{\extracolsep{\fill}}cccc}
\hline
\hline
State & $\Xi_b \bar K$ & $\Xi'_b \bar K$ & Total  \\
\hline
$6405.2$ & $0.024$ & - & $0.024$  \\
 $6427$ & - & - &- \\
 $6465 $ & $1.67$ & $0.755$  & $2.42$\\
 $6508$ & - & -  & -\\
 \hline\hline
\end{tabular*}
\end{table*}
  We can see that within our model, that accounts for the main components in the coupled channels,
  the width of the $6427$ and $6508$ states is zero.
  Comparing that with the largest width that we obtain of $2.42$ MeV,
  we can reasonably expect that the widths of these states should be smaller than $1$ MeV.
  The $6405$ state only decays into $\Xi_b \bar K$,
  and since the coupling of the state to this channel is very small,
  the width that we obtain is a mere $0.024$ MeV.
  On the other hand, we can split the total width of the $6465$ state into two partial decays,
  one with $1.67$ MeV to the  $\Xi_b \bar K$ channel and another one with $0.75$ MeV into the $\Xi'_b \bar K$  channel.
  This is valuable information that could be contrasted with experiment in future runs.
  The widths, the partial decay widths, and the spin and parity of these possible states would be very different from any quark model prediction.
  By comparison, the widths evaluated with the $^3P_0$ models from the quark model states predict widths of the order of $15$ MeV for $2S$ states
  with masses $6483$ and $6495$ MeV in Ref.~\cite{21chen}.
  In Ref.~\cite{Qifang}, the $2S$ state of $\frac{1}{2}^+$ and $\frac{3}{2}^+$ have widths around $50$ MeV.
  On the other hand, in Ref.~\cite{18chen} the widths obtained for two $2S$ states are smaller than $2$ MeV.
  With these different results in different quark models, it is difficult to assert the values of the width for the $2S$ states,
  but they all share that the parity is positive in contrast with the negative parity of the states that we predict in that region.
  Our predictions for the widths in Table \ref{tab:new} are quite stable, and it would be most interesting
  to see if future runs of the experiments can measure these rates.

\section{Concluding remarks}
We showed that the results obtained in Ref.~\cite{Omegab} for molecular $\Omega_b$ states from the meson-baryon interaction in coupled channels
constructed by analogy to the $\Omega_c$ states of Ref.~\cite{Omegac}, which agree with the experimentally observed states \cite{ExpOmegac},
give rise to four states with energies in remarkable agreement with some peaks
that can be observed in the high energy part of the $\Omega_b$ experimental spectrum~\cite{LHCb}.
While the statistical significance is poor, we noticed that it is about the same as for the first of the $\Omega_b$ states claimed in Ref.~\cite{LHCb}.
While we refrained from claiming that these peaks correspond to $\Omega_b$ states,
the agreement found is reason enough to suggest increasing the statistics and looking in detail in the region
where these peaks have appeared to hopefully find an answer to the suggestion made in this work.

We also made a comparison of our results with different quark models,
concluding that the recently observed $\Omega_b$ states could well be ordinary $1P$ excitations of three quark systems,
while the states that we predicted cannot accommodate these states.
On the other hand, the quark models predicted states around $6450$ MeV,
where we found our $\frac{1}{2}^-, \frac{3}{2}^-$ molecular states, but as $2S$ excitations, which have positive parity.
We also evaluated partial decay widths to different channels,
and while there were discrepancies in the widths of these states predicted by different quark models,
it was clear that the predictions were quite different from the results that we obtained.
It is thus clear that precise measurements in that region,
with enough statistics to claim new states and determine the widths, partial decay widths,
and eventually the spin and parity of the states,
could clearly discriminate between the excited conventional three quark states and the molecular states that we proposed.

\section{ACKNOWLEDGEMENTS}
This work is partly supported by the National Natural Science Foundation of China under Grants No. 11975083, No. 11947413,
No. 11847317 and No. 11565007. This work is also partly supported by the Spanish Ministerio de
Economia y Competitividad and European FEDER funds under Contracts No. FIS2017-
84038-C2-1-P B and No. FIS2017-84038-C2-2-P B, the Generalitat Valenciana in the
program Prometeo II-2014/068, and the project Severo Ochoa of IFIC, SEV-2014-0398.


\end{document}